\documentclass[journal]{IEEEtran}

\ifCLASSINFOpdf
\else
   \usepackage[dvips]{graphicx}
\fi
\usepackage{url}
\usepackage{amsmath,amssymb,amsfonts}
\usepackage{graphicx}
\usepackage{multirow}

\begin{document}

\title{\v{C}ech Complex Generation with Homotopy Equivalence Framework for Myocardial Infarction Diagnosis using Electrocardiogram Signals}

\author{Srikireddy Dhanunjay Reddy, Pujayita Deb, Tharun Kumar Reddy Bollu

\vspace{-9mm}
}
\maketitle
\begin{abstract}
Early and optimal identification of cardiac anomalies, especially Myocardial infarction (MCI) can aid the individual in obtaining prompt medical attention to mitigate the severity. Electrocardiogram (ECG) is a simple non-invasive physiological signal modality, that can be used to examine the electrical activity of heart tissue. Existing methods for MCI detection mostly rely on the temporal, frequency, and spatial domain analysis of the ECG signals. These conventional techniques lack in effective identification of cardiac cycle inter-dependency during diagnosis. Hence, there is an emerging need for incorporating the underlying connectivity of the intra-sessional cardiac cycles for improved anomaly detection. This article proposes a novel framework for ECG signal analysis and classification using persistent homological features through \v{C}ech Complex generation with homotopy equivalence check, by taking the above-mentioned emerging needs into account. Homological features like persistent birth-death rates, betti curves, and persistent entropy provide transparency of the regional and cardiac cycle connectivity when combined with Machine Learning (ML) models. The proposed framework is assessed using publicly available datasets (MIT-BIH and PTB), and the performance metrics of machine learning models indicate its efficacy in classifying Normal Sinus Rhythm (NSR), MCI, and non-MCI subjects, achieving a 2.8\% mean improvement in AUC (area under the ROC curve) over existing approaches. 
\end{abstract}
\IEEEpeerreviewmaketitle

\vspace{-2mm}
\section{Introduction}
ECG is a wearable non-invasive technique for monitoring and detecting cardiac anomalies through the electrical activity of heart tissue \cite{i1} \cite{i2}. MCI is one of the most challenging cardiovascular diseases (CVD) leading to heart tissue failures and cardiac arrests \cite{i1}. Timely automated prediction of the MCI condition can help health practitioners in reducing the mortality rate due to MCI severity \cite{i3}. The presence of artifacts and the non-stationarity nature of ECG signal demands prominent feature extraction and robust anomaly detection. Early works used various artifact removal and feature extraction methods prior to classification \cite{i4}-\cite{i0}
. Application of ML models and deep learning (DL) frameworks using raw ECG signals and its extracted features became the traditional and effective combination for CVD prediction \cite{i12} \cite{i13}.\\
Topological Data Analysis (TDA) is an emerging technique, to analyze the connectivity and complexity of multi-channel physiological signals \cite{i14}. Persistent homological features of homological groups using simplicial complex generation serves as robust biomarkers in finding inter-trial connectivity \cite{i15} \cite{i16}. Distribution and equivalence of point cloud data points affect simplicial complex evolution and obtained features. Homotopy equivalence verification of the generated complex helps reduce outlier effects on other data points \cite{i17}. The retrieved features can be provided to the ML models for the targeted task of anomaly detection.\\
Feature Extraction is a crucial step in recognizing the underlying connectivity patterns of the signal data. In contrast, the homological features derived during the formation of the simplicial complex illustrate the interconnection among the data points in the point cloud, which includes the subject's trial data. This letter analyzes the efficiency of the proposed \v{C}ech complex framework in anomaly prediction, relative to existing automated CVD prediction approaches. The primary contributions to this effect are as follows: 
\begin{itemize}
    \item To the best of our knowledge, through this article, a novel \v{C}ech complex framework is proposed to classify the ECG signals by an analysis of the connectivity among intra-sessional trials using persistent homological features.
    \item Homotopy equivalence checking condition is proposed to nullify the impact of the outliers in point clouds, without removing its significance in \v{C}ech complex generation.
\end{itemize}
Furthermore, the methods that have been proposed are explored in Section II. Within Section III, a description of the datasets is provided. Finally, the results and conclusion are discussed in Sections IV and V.
\vspace{-5mm}
\section{Proposed Method}
\subsection{\v{C}ech Complex Generation}
\begin{figure}[htbp]
\centering
\centerline{ \includegraphics[scale=0.7]{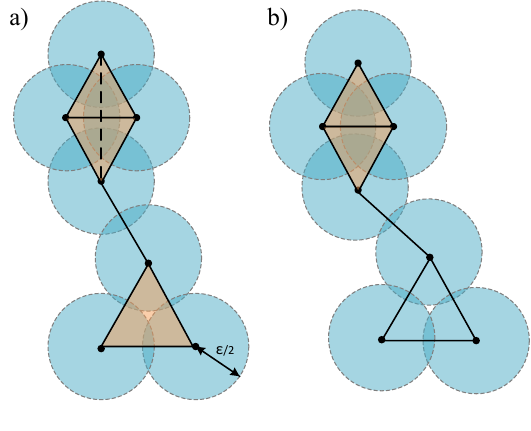}}
\vspace{-5mm}
\caption{a) Vietoris-Rips and b) \v{C}ech complexes comparison of non-empty intersection}
\label{VRC comparison}
\end{figure}
\begin{figure*}[htbp]
\centering
\centerline{ \includegraphics[width=\textwidth]{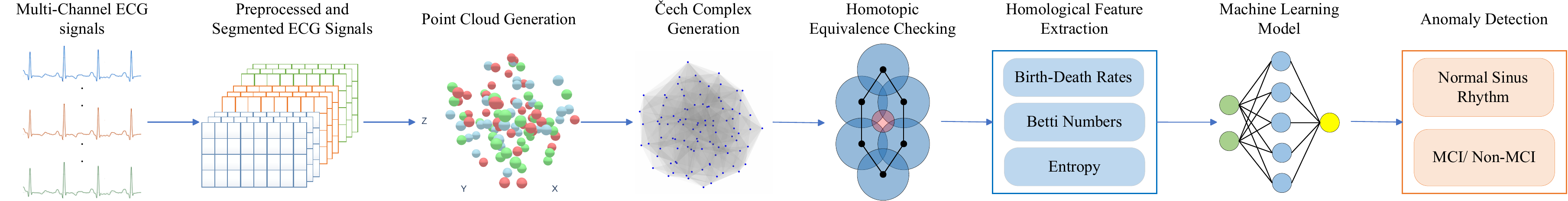}}
\caption{\v{C}ech complex framework for multi-channel ECG signal classification with homotopic equivalence check}
\label{framework}
\vspace{-2mm}
\end{figure*}
Consider a point cloud $P=\{p_1, p_2, p_3,\cdots, p_k\}$, representing $K$ finite set of data points in $d$-dimensional metric or topological space, where each $p_i \in$ $\mathbb{R}^d$. For the data points $p_i$ and $p_j$, $\delta (p_i,p_j)$ denotes the distance between these points.
\begin{equation}
    \delta(p_i,p_j)=||p_i-p_j||^2= \sqrt{\sum_{k=1}^K(p_i^k-p_j^2)^2}
\end{equation}
To generate \v{C}ech complex \v{C}$_{\epsilon}(P)$ for the given $P$, an open ball $B_{\epsilon}(p_i)$ of diameter $\epsilon$ is constructed around each $p_i$. Where $B_{\epsilon}(p_i)$ is defined as:
\begin{equation}
    B_{\epsilon}(p_i)={x \in \mathbb{R}^d; \delta(p_i,x)= \frac{\epsilon}{2}}
\end{equation}
$x$ denotes any points on the ball circumference. \v{C}$_{\epsilon}(P)$ at scale $\frac{\epsilon}{2}$ is constructed by looking at intersection of $B_{\epsilon}(p_i); \forall i$. 0-simplex (point), 1-simplex (edge), 2-simplex (triangle), 3-simplex (tetrahedron), and their higher-order simplices are the building blocks of \v{C}$_{\epsilon}(P)$. 0-simplex is formed by each $p_i$ in $P$. 1-simplex is formed between two points $p_i$ and $p_j$, if the boundaries of  $B_{\epsilon}(p)$ intersects.
\begin{equation}
    B_{\epsilon}(p_i) \cap B_{\epsilon}(p_j) \neq \phi , \iff \delta(p_i,p_j) \leq \epsilon
\end{equation}
Similarly, to form the $K-1$ simplex, corresponding $B_{\epsilon}(p)$ has to follow the non-empty intersection condition as follows:
\begin{equation}
    \bigcap_{k=1}^K B_{\epsilon}(p_k) \neq \phi, \iff \delta(P)\leq\epsilon 
\end{equation}
Formally, \v{C}$_\epsilon(P)$ at scale $\frac{\epsilon}{2}$ is the set of simplices $\sigma$ formed by the point cloud $P$, that can be denoted as:
\begin{equation}
    \check{C}_{\epsilon}(P)=\{\sigma \subset P : \bigcap_{P \in \sigma} B_{\epsilon}(P)\neq \phi\}
\end{equation}
$\check{C}_{\epsilon}(P)$ can be generated with the evolution of the  $B_{\epsilon}(p_i) ; 0 \rightarrow \epsilon $. considering the evolution of the $\check{C}_{\epsilon}(P)$ with different ball radii set $\{\epsilon_1, \epsilon_2, \epsilon_3, \cdot, \epsilon_m\}$, the relation among the complexes at different $\epsilon$ values are given as follows:
\begin{equation}
    \check{C}_{\epsilon_1}(P) \subseteq\check{C}_{\epsilon_2}(P) \subseteq\check{C}_{\epsilon_3}(P) \subseteq . . . \check{C}_{\epsilon_m}(P) 
\end{equation}
With the condition being $\epsilon_1 < \epsilon_2< \epsilon_3< ... < \epsilon_m.$ Figure \ref{VRC comparison} shows the example of generated Vietoris-Rips and \v{C}ech complexes comparison with limited data points. The generation strategy of the multi-channel physiological signal is discussed in  \cite{p1} (section 2.2). From figure \ref{VRC comparison}, it can be observed that the Vietoris-Rips complex has more homological groups when compared to \v{C}ech complex. Additional 2-simplex and 3-simplex are formed in the Vietoris-Rips complex without satisfying the requirement of a non-empty intersection, which may result in homotopy inequivalence. The corresponding data points meet the distance metric condition but fail to satisfy the non-empty condition in (4). Subsequent sections will address the significance and applications of homotopy equivalence.
\vspace{-4mm}
\subsection{Homotopy Equivalence Check}
\begin{figure}[htbp]
\centering
\centerline{ \includegraphics[width=\linewidth]{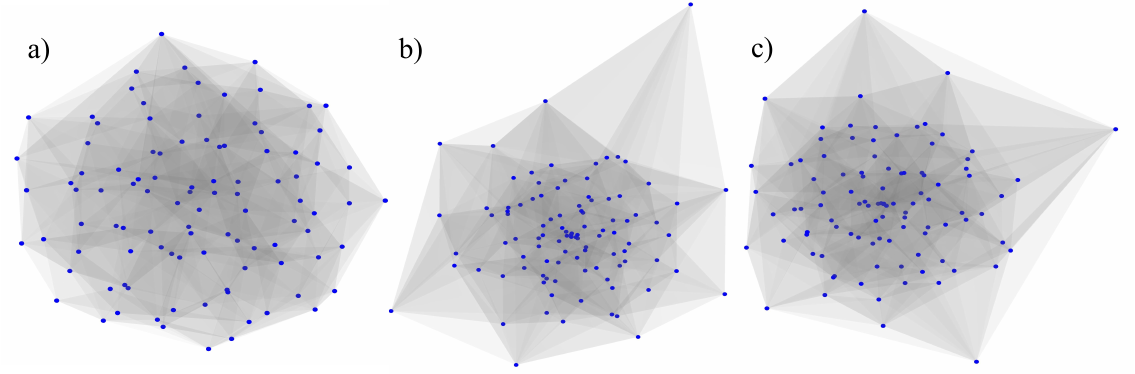}}
\vspace{-3mm}
\caption{Generated \v{C}ech complexes for a) NSR, b) MCI, and c) non-MCI subjects}
\label{cech complex}
\end{figure}
Generated \v{C}ech complex is said to be homotopically equivalent if it can transformed into the union of $B_\epsilon (P)$ without tearing or gluing using new points. Consider mapping functions as shown below:
\begin{equation}
    f: \check{C}_\epsilon(P) \rightarrow \bigcup_{k=1}^K B_\epsilon(p_k)
\end{equation}
\begin{equation}
    g: \bigcup_{k=1}^K B_\epsilon(p_k) \rightarrow \check{C}_\epsilon(P)
\end{equation}
Then, mapping functions $f$ and $g$ satisfies the equivalent condition:
\begin{equation}
    f \circ g \simeq g \circ f \simeq id_P
\end{equation}
'$\circ$' represents the composition of two maps, and '$id_P$' represents identical mapping on topological space. In addition to this mapping condition, if \v{C}$_\epsilon(P)$ satisfies the following non-empty contradiction condition, then the generated \v{C}ech complex is homotopically equivalent.
\begin{equation}
     \bigcap_{k=1}^{K} B_{\epsilon}(p_i)\neq \phi\
\end{equation}
Figure \ref{cech complex} shows \v{C}$_\epsilon(P)$ of different classes of subjects that exist in datasets used in this article. It has been noted that $P$ associated with MCI and non-MCI exhibits outliers, indicating an unequal distribution of data points. The homotopy equivalence criterion will maintain the significance of these data points, and their effects can be mitigated by eliminating the transient simplex feature data. 
\vspace{-3mm}
\subsection{Persistent Homological Features} \label{homological feature section}
Homological features describe the evolution of the \v{C}ech complex with various $\epsilon$ values. In this process, homologial groups, connected components $(H_0)$, loops $(H_1)$, and voids $(H_2)$ are generated. Connected simplices are the building blocks of these homological groups. Features like birth-death rates $(b,d)$, betti curves $\beta_i(\epsilon)$, and entropy of simplices $\mathrm{E}$ are used as the measure on connectivity among trials. Expression of these features concerning $(b,d)$ can be noted as follows:
\begin{equation}
    f(b,d)=\{(b,d]:b,d \in \mathbb{R}\} \cup \{{b,\infty}: b \in \mathbb{R}\}
\end{equation}
\begin{equation}
    \beta_k(\epsilon)= \sum_{k=1}^{K}w(b-d) \overrightarrow{1}_k
\end{equation}
\begin{equation}
    \mathrm{E}=\sum_{k=1}^{K} \frac{d_k-b_k}{L} log(\frac{d_k-b_k}{L}); L=\sum_{k=1}^{K}(d_k-b_k)
\end{equation}
$f(b,d)$ represents the linear representation function of birth-death of the homological groups. $\beta_k(\epsilon)$ represents the population of the different homological groups with the evolution of $\epsilon$ in terms of increasing monotonic function $w(.)$ and 1's vector. $\mathrm{E}$ explains the underlying connectivity of subject-specific sessions concerning homological group evolution. In general, outliers have minimal (b,d) values based on the distance metric. Focusing on groups that persist across a wide range of $\epsilon$ evolution, eliminates the shortest living homological features from homotopy equivalent verified \v{C}ech complex. Moreover, extracted homological features can be provided to ML models for anomaly prediction. 
\vspace{-3mm}
\subsection{Proposed \v{C}ech Complex Framework}
Consider ECG signal data $X \in \mathbb{R}^{ch \times s}$ with $ch$ number of channels and $s$ number of time samples per channel. Initially, $X$ is subjected to a two-stage median filter and Discrete Wavelet Transform filtering techniques \cite{i3}, to remove the myographic and baseline wander contamination from the cardiac signal data. Further, $X$ is segmented into 4-second trials and then vectorized to get $X \in \mathbb{R}^{n \times t}$, n represents the number of trials, and t represents the number of time samples per trial. Now, $X$ is decomposed into the orthonormal vectors $U$ and the eigenvalue matrix $\Lambda$ as follows:
\begin{equation}
    X_{n \times t}= U_{n \times n} \Lambda_{n \times t} U^T_{n \times n}
\end{equation}
Proposed \v{C}ech complex framework generated a three-dimensional point cloud $P \in \mathbb{R}^{n \times 3}$ by selecting the 3 high-ranked eigenvalues (to observe upto third-order homological group distribution). $d$-dimensional point cloud can be generated for different applications, by mapping the $d$-eigenvalues of $\Lambda_k$ onto the appropriate $U_d$ eigenvectors as follows:
\begin{equation}
    P=U_k \Lambda^{\frac{1}{2}}_k
\end{equation}
Further, $P$ is utilized to generate $\check{C}_\epsilon(P)$ under homotopy equivalent condition (10). Persistent homological features like $(b,d)$, $\beta_k(\epsilon)$, and $\mathrm{E}$ of simplices are calculated to do anomaly detection using various ML models. Figure \ref{framework}, indicates the overall flow of the proposed \v{C}ech complex framework.
\begin{table}[]
\centering
\caption{$\mathrm{E}_\mu$ and $\mathrm{E}_{\sigma^2}$ comparison of NSR, MCI, and non-MCI for all homological groups}
\vspace{-2mm}
\label{entropy comparison}
\resizebox{\columnwidth}{!}{%
\begin{tabular}{|c|ccc|ccc|ccc|}
\hline
\multirow{2}{*}{\textbf{$\mathrm{E}$}} & \multicolumn{3}{c|}{\textbf{NSR}}                            & \multicolumn{3}{c|}{\textbf{MCI}}                            & \multicolumn{3}{c|}{\textbf{non-MCI}}                       \\ \cline{2-10} 
                            & \multicolumn{1}{c|}{H0}   & \multicolumn{1}{c|}{H1}   & H2   & \multicolumn{1}{c|}{H0}   & \multicolumn{1}{c|}{H1}   & H2   & \multicolumn{1}{c|}{H0}  & \multicolumn{1}{c|}{H1}   & H2   \\ \hline
\textbf{$\mathrm{E}_\mu$}                 & \multicolumn{1}{c|}{0.35} & \multicolumn{1}{c|}{0.28} & 0.32 & \multicolumn{1}{c|}{0.55} & \multicolumn{1}{c|}{0.6}  & 0.57 & \multicolumn{1}{c|}{0.5} & \multicolumn{1}{c|}{0.53} & 0.51 \\ \hline
\textbf{$\mathrm{E}_{\sigma^2}$}                 & \multicolumn{1}{c|}{0.05} & \multicolumn{1}{c|}{0.04} & 0.02 & \multicolumn{1}{c|}{0.12} & \multicolumn{1}{c|}{0.14} & 0.16 & \multicolumn{1}{c|}{0.1} & \multicolumn{1}{c|}{0.11} & 0.09 \\ \hline
\end{tabular}%
}
\vspace{-4mm}
\end{table}
\begin{figure*}[htbp]
\centering
\centerline{ \includegraphics[scale=0.56]{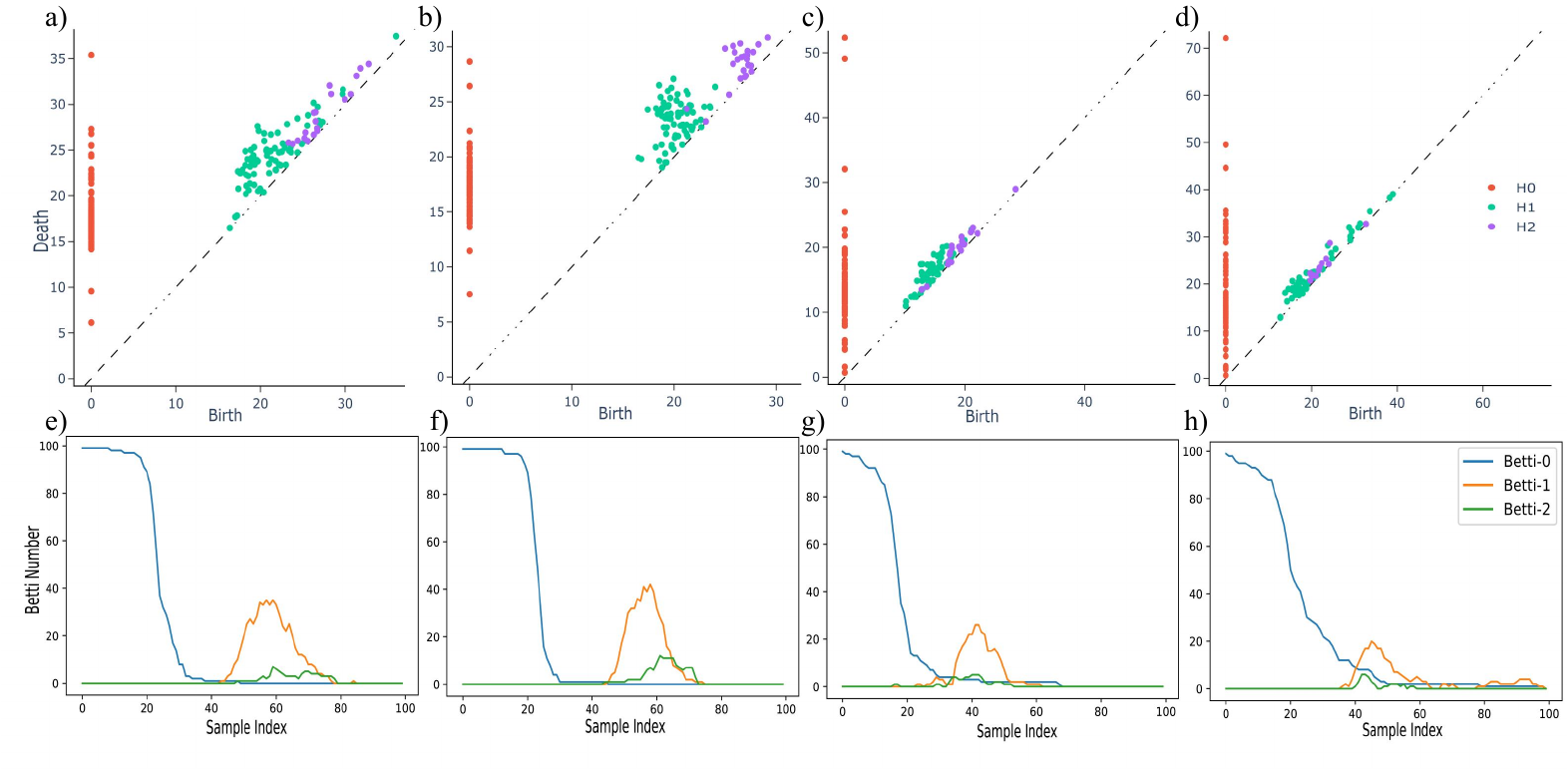}}
\vspace{-3mm}
\caption{Homological features comparison; UP: a-b) NSR, c) MCI, and d) non-MCI subject persistent homological data plots; DOWN: e-f) NSR, g) MCI, and h) non-MCI subject betti curve plots }
\label{results 1}
\vspace{-3.5mm}
\end{figure*}
\begin{table*}[]
\centering
\renewcommand{\arraystretch}{1.2} 
\normalsize 
\caption{Performance comparison of different models on PTB \cite{d1} and MITBIH-AD \cite{d2} datasets using \v{C}ech complex framework}
\vspace{-2mm}
\label{Model performance comparison}
\resizebox{\linewidth}{!}{%
\begin{tabular}{clcccccccccccccccccccccccc}
\hline
\multicolumn{2}{|c|}{\multirow{3}{*}{\textbf{Classifiers}}} & \multicolumn{8}{c|}{\textbf{NSR   Vs MCI}} & \multicolumn{8}{c|}{\textbf{MCI   Vs non-MCI}} & \multicolumn{8}{c|}{\textbf{NSR   Vs MCI Vs non-MCI}} \\ \cline{3-26} 
\multicolumn{2}{|c|}{} & \multicolumn{2}{c|}{\textbf{Acc±SD}} & \multicolumn{2}{c|}{\textbf{AUC}} & \multicolumn{2}{c|}{\textbf{f1 score}} & \multicolumn{2}{c|}{\textbf{Kappa}} & \multicolumn{2}{c|}{\textbf{Acc±SD}} & \multicolumn{2}{c|}{\textbf{AUC}} & \multicolumn{2}{c|}{\textbf{f1 score}} & \multicolumn{2}{c|}{\textbf{Kappa}} & \multicolumn{2}{c|}{\textbf{Acc±SD}} & \multicolumn{2}{c|}{\textbf{AUC}} & \multicolumn{2}{c|}{\textbf{f1 score}} & \multicolumn{2}{c|}{\textbf{Kappa}} \\ \cline{3-26} 
\multicolumn{2}{|c|}{} & \multicolumn{1}{c|}{\textbf{I}} & \multicolumn{1}{c|}{\textbf{II}} & \multicolumn{1}{c|}{\textbf{I}} & \multicolumn{1}{c|}{\textbf{II}} & \multicolumn{1}{c|}{\textbf{I}} & \multicolumn{1}{c|}{\textbf{II}} & \multicolumn{1}{c|}{\textbf{I}} & \multicolumn{1}{c|}{\textbf{II}} & \multicolumn{1}{c|}{\textbf{I}} & \multicolumn{1}{c|}{\textbf{II}} & \multicolumn{1}{c|}{\textbf{I}} & \multicolumn{1}{c|}{\textbf{II}} & \multicolumn{1}{c|}{\textbf{I}} & \multicolumn{1}{c|}{\textbf{II}} & \multicolumn{1}{c|}{\textbf{I}} & \multicolumn{1}{c|}{\textbf{II}} & \multicolumn{1}{c|}{\textbf{I}} & \multicolumn{1}{c|}{\textbf{II}} & \multicolumn{1}{c|}{\textbf{I}} & \multicolumn{1}{c|}{\textbf{II}} & \multicolumn{1}{c|}{\textbf{I}} & \multicolumn{1}{c|}{\textbf{II}} & \multicolumn{1}{c|}{\textbf{I}} & \multicolumn{1}{c|}{\textbf{II}} \\ \hline
\multicolumn{2}{|c|}{Random Forest} & \multicolumn{1}{c|}{98.0±1.58} & \multicolumn{1}{c|}{95.2±2.1} & \multicolumn{1}{c|}{0.99} & \multicolumn{1}{c|}{0.98} & \multicolumn{1}{c|}{0.98} & \multicolumn{1}{c|}{0.96} & \multicolumn{1}{c|}{0.94} & \multicolumn{1}{c|}{0.94} & \multicolumn{1}{c|}{96.8±2.0} & \multicolumn{1}{c|}{90.1±2.3} & \multicolumn{1}{c|}{0.98} & \multicolumn{1}{c|}{0.94} & \multicolumn{1}{c|}{0.98} & \multicolumn{1}{c|}{0.91} & \multicolumn{1}{c|}{0.94} & \multicolumn{1}{c|}{0.87} & \multicolumn{1}{c|}{89.7±2.5} & \multicolumn{1}{c|}{85.3±2.9} & \multicolumn{1}{c|}{0.94} & \multicolumn{1}{c|}{0.9} & \multicolumn{1}{c|}{0.93} & \multicolumn{1}{c|}{0.86} & \multicolumn{1}{c|}{0.92} & \multicolumn{1}{c|}{0.82} \\ \hline
\multicolumn{2}{|c|}{SVM} & \multicolumn{1}{c|}{95.6±2.3} & \multicolumn{1}{c|}{92.8±2.5} & \multicolumn{1}{c|}{0.94} & \multicolumn{1}{c|}{0.96} & \multicolumn{1}{c|}{0.93} & \multicolumn{1}{c|}{0.94} & \multicolumn{1}{c|}{0.9} & \multicolumn{1}{c|}{0.91} & \multicolumn{1}{c|}{91.5±2.5} & \multicolumn{1}{c|}{88.3±2.6} & \multicolumn{1}{c|}{0.88} & \multicolumn{1}{c|}{0.92} & \multicolumn{1}{c|}{0.86} & \multicolumn{1}{c|}{0.89} & \multicolumn{1}{c|}{0.83} & \multicolumn{1}{c|}{0.85} & \multicolumn{1}{c|}{86.7±2.7} & \multicolumn{1}{c|}{83.9±2.9} & \multicolumn{1}{c|}{0.85} & \multicolumn{1}{c|}{0.88} & \multicolumn{1}{c|}{0.84} & \multicolumn{1}{c|}{0.85} & \multicolumn{1}{c|}{0.8} & \multicolumn{1}{c|}{0.8} \\ \hline
\multicolumn{2}{|c|}{MLP} & \multicolumn{1}{c|}{96.3±2.0} & \multicolumn{1}{c|}{93.5±2.4} & \multicolumn{1}{c|}{0.95} & \multicolumn{1}{c|}{0.97} & \multicolumn{1}{c|}{0.94} & \multicolumn{1}{c|}{0.95} & \multicolumn{1}{c|}{0.92} & \multicolumn{1}{c|}{0.92} & \multicolumn{1}{c|}{94.7±2.3} & \multicolumn{1}{c|}{89.4±2.5} & \multicolumn{1}{c|}{0.92} & \multicolumn{1}{c|}{0.93} & \multicolumn{1}{c|}{0.87} & \multicolumn{1}{c|}{0.9} & \multicolumn{1}{c|}{0.88} & \multicolumn{1}{c|}{0.88} & \multicolumn{1}{c|}{87.3±2.4} & \multicolumn{1}{c|}{84.8±2.7} & \multicolumn{1}{c|}{0.91} & \multicolumn{1}{c|}{0.89} & \multicolumn{1}{c|}{0.89} & \multicolumn{1}{c|}{0.86} & \multicolumn{1}{c|}{0.88} & \multicolumn{1}{c|}{0.81} \\ \hline
\multicolumn{2}{|c|}{Logistic Regression} & \multicolumn{1}{c|}{94.5±2.5} & \multicolumn{1}{c|}{91.7±2.9} & \multicolumn{1}{c|}{0.93} & \multicolumn{1}{c|}{0.95} & \multicolumn{1}{c|}{0.92} & \multicolumn{1}{c|}{0.93} & \multicolumn{1}{c|}{0.89} & \multicolumn{1}{c|}{0.9} & \multicolumn{1}{c|}{89.3±3.0} & \multicolumn{1}{c|}{86.7±3.1} & \multicolumn{1}{c|}{0.87} & \multicolumn{1}{c|}{0.91} & \multicolumn{1}{c|}{0.85} & \multicolumn{1}{c|}{0.88} & \multicolumn{1}{c|}{0.84} & \multicolumn{1}{c|}{0.86} & \multicolumn{1}{c|}{85.1±2.9} & \multicolumn{1}{c|}{82.6±3.0} & \multicolumn{1}{c|}{0.84} & \multicolumn{1}{c|}{0.87} & \multicolumn{1}{c|}{0.82} & \multicolumn{1}{c|}{0.84} & \multicolumn{1}{c|}{0.78} & \multicolumn{1}{c|}{0.79} \\ \hline
\multicolumn{2}{|c|}{Decision Tree} & \multicolumn{1}{c|}{93.1±3.1} & \multicolumn{1}{c|}{90.4±3.2} & \multicolumn{1}{c|}{0.91} & \multicolumn{1}{c|}{0.94} & \multicolumn{1}{c|}{0.9} & \multicolumn{1}{c|}{0.92} & \multicolumn{1}{c|}{0.87} & \multicolumn{1}{c|}{0.88} & \multicolumn{1}{c|}{90.5±3.2} & \multicolumn{1}{c|}{85.2±3.4} & \multicolumn{1}{c|}{0.85} & \multicolumn{1}{c|}{0.89} & \multicolumn{1}{c|}{0.83} & \multicolumn{1}{c|}{0.86} & \multicolumn{1}{c|}{0.86} & \multicolumn{1}{c|}{0.81} & \multicolumn{1}{c|}{83.2±3.4} & \multicolumn{1}{c|}{80.9±3.5} & \multicolumn{1}{c|}{0.82} & \multicolumn{1}{c|}{0.84} & \multicolumn{1}{c|}{0.8} & \multicolumn{1}{c|}{0.83} & \multicolumn{1}{c|}{0.75} & \multicolumn{1}{c|}{0.78} \\ \hline
\multicolumn{25}{l}{* (ACC±SD)   represents Accuracy ± Standard Deviation) value in a percentage format.  I and II refers to the datasets {[}18{]} →   12-leads and {[}19{]} → 2-leads} & \multicolumn{1}{l}{}
\end{tabular}%
}
\vspace{-4mm}
\end{table*}
\begin{table}[]
\centering
\caption{Comparison of the proposed features performance on PTB dataset with existing methods}
\vspace{-2mm}
\label{feature performance comparison}
\resizebox{\columnwidth}{!}{%
\begin{tabular}{ccccc}
\hline
\multicolumn{1}{|c|}{\textbf{Approach}} & \multicolumn{1}{c|}{\textbf{Complexity}} & \multicolumn{1}{c|}{\textbf{Classification}} & \multicolumn{1}{c|}{\textbf{F1 score}} & \multicolumn{1}{c|}{\textbf{AUC}} \\ \hline
\multicolumn{1}{|c|}{U-net \cite{r1}} & \multicolumn{1}{c|}{$O(n^2m)$} & \multicolumn{1}{c|}{NSR Vs MCI} & \multicolumn{1}{c|}{92.53} & \multicolumn{1}{c|}{97.96} \\ \hline
\multicolumn{1}{|c|}{CNN \cite{r2}} & \multicolumn{1}{c|}{$O(n^2md)$} & \multicolumn{1}{c|}{NSR Vs MCI} & \multicolumn{1}{c|}{97.8} & \multicolumn{1}{c|}{96.54} \\ \hline
\multicolumn{1}{|c|}{ECG-SMART-NET \cite{r3}} & \multicolumn{1}{c|}{$O(n^2md)$} & \multicolumn{1}{c|}{NSR Vs MCI} & \multicolumn{1}{c|}{64.2} & \multicolumn{1}{c|}{92.35} \\ \hline
\multicolumn{1}{|c|}{\multirow{2}{*}{RNN \cite{r4}}} & \multicolumn{1}{c|}{\multirow{2}{*}{$O(nm^2h)$}} & \multicolumn{1}{c|}{NSR Vs MCI} & \multicolumn{1}{c|}{96.24} & \multicolumn{1}{c|}{97.8} \\ \cline{3-5} 
\multicolumn{1}{|c|}{} & \multicolumn{1}{c|}{} & \multicolumn{1}{c|}{MCI Vs non-MCI} & \multicolumn{1}{c|}{94.91} & \multicolumn{1}{c|}{96.9} \\ \hline
\multicolumn{1}{|c|}{\multirow{2}{*}{RNN \cite{r5}}} & \multicolumn{1}{c|}{\multirow{2}{*}{$O(nm^2h)$}} & \multicolumn{1}{c|}{NSR Vs MCI} & \multicolumn{1}{c|}{97.65} & \multicolumn{1}{c|}{98} \\ \cline{3-5} 
\multicolumn{1}{|c|}{} & \multicolumn{1}{c|}{} & \multicolumn{1}{c|}{MCI Vs non-MCI} & \multicolumn{1}{c|}{96.72} & \multicolumn{1}{c|}{97.24} \\ \hline
\multicolumn{1}{|c|}{Wavelet Transform \cite{r6}} & \multicolumn{1}{c|}{$O(n^2\log n)$} & \multicolumn{1}{c|}{NSR Vs MCI} & \multicolumn{1}{c|}{97.14} & \multicolumn{1}{c|}{97.86} \\ \hline
\multicolumn{1}{|c|}{Time-Frequency \cite{r7}} & \multicolumn{1}{c|}{$O(n^2m)$} & \multicolumn{1}{c|}{NSR Vs MCI} & \multicolumn{1}{c|}{92.53} & \multicolumn{1}{c|}{93.26} \\ \hline
\multicolumn{1}{|c|}{\multirow{3}{*}{\begin{tabular}[c]{@{}c@{}}Homological Features\\ (proposed)\end{tabular}}} & \multicolumn{1}{c|}{\multirow{3}{*}{$O(n^2m)$}} & \multicolumn{1}{c|}{NSR Vs MCI} & \multicolumn{1}{c|}{98.97} & \multicolumn{1}{c|}{99.2} \\ \cline{3-5} 
\multicolumn{1}{|c|}{} & \multicolumn{1}{c|}{} & \multicolumn{1}{c|}{MCI Vs non-MCI} & \multicolumn{1}{c|}{98.67} & \multicolumn{1}{c|}{98.81} \\ \cline{3-5} 
\multicolumn{1}{|c|}{} & \multicolumn{1}{c|}{} & \multicolumn{1}{c|}{NSR Vs MCI Vs non-MCI} & \multicolumn{1}{c|}{93.12} & \multicolumn{1}{c|}{94.26} \\ \hline
\multicolumn{5}{l}{* $n$: no. of trials, $m$: trial length, $d$: model depth, $h$: no. of hidden layers}
\end{tabular}%
}
\vspace{-7mm}
\end{table}

\section{Dataset Description}
The proposed \v{C}ech complex framework is implemented and validated on the PTB Diagnostic ECG Database (PTB) \cite{d1} and MIT-BIH Arrhythmia Database (MIT-BIH AD) \cite{d2}. PTB comprises a 2-minute, 15-lead high-resolution ECG signal data, including 52 NSR, 148 MCI, and 68 non-MCI subject samples, recorded at a sampling frequency of 1000 Hz. In this work, 12 lead ECG data is utilized to validate the proposed framework. Further, ECG signals are processed to eliminate noise and baseline wander before being segmented into 4-second intervals and then homological features are extracted.\\  
MIT-BIH AD comprises a 30-minute, 2-lead (V5 and II) 11-bit resolution ECG recording sampled at 360 Hz from 48 participants, featuring labeled sessions of different arrhythmic conditions. In this article, arrhythmic conditions that may lead to Mild Cognitive Impairment (MCI), such as ventricular fibrillation, ventricular tachycardia, atrial fibrillation, and sinus bradycardia, are labeled as MCI. Conversely, conditions including atrial flutter, supraventricular tachycardia, bigeminy, and trigeminy are classified as non-MCI. Further preprocessed and 4-sec segmented trials are passed to the proposed framework.
\vspace{-4mm}
\section{Results \& Discussion}
\subsection{Homological feature analysis}
Homological features are caluclated from the generated \v{C}ech complex as discussed in the section \ref{homological feature section}. Based on the trial wise (b,d) rate of each homological groups $(H_0, H_1, H_2)$, entropy $\mathrm{E}$ of a subject is calculated. Further, mean entropy $(\mathrm{E}_\mu)$ and entropy variance $(\mathrm{E}_{\sigma^2})$ of each class homological group is calculated as shown in Table \ref{entropy comparison}. Increased $\mathrm{E}_\mu$ and $\mathrm{E}_{\sigma^2}$ of MCI and non-MCI classes represents the high non-stationarity in comparison with NSR class. In Figure \ref{results 1}, homological features from MSR, MCI, and non-MCI subjects are visualized. (UP) shows the persistent homological plot of all homological groups (b,d) values. (DOWN) shows the Betti curves $\beta_k(\epsilon)$ demonstrating group evolution (b,d). Due to their complex and significant non-stationarities, MCI and non-MCI features overlap more than NSR. The large betti curve distribution for NSR indicates its stable and simpler point cloud structure. Naive MCI and non-MCI distribution imply unstable and complex structures.
\vspace{-4mm}
\subsection{Myocardial Infraction detection}
Extracted homological features from the proposed \v{C}ech complex framework are fed to diverse ML models with a 5-fold cross-validation scheme by the 80:20 train-test split strategy. Parameter configuration of the best-performed model details are discussed as follows:\\
\emph{Random Forest}: Random Forest classifier was optimized with 275 trees and a max\_depth of 21 to balance complexity and performance. The model used a 4 min\_samples\_split and 3 min\_samples\_leaf to control overfitting.\\
\emph{SVM}: Support Vector Machine (SVM) classifier achieved optimal performance using a Radial Basis Function (RBF) kernel and a regularization parameter (C) of 0.5. The gamma parameter was set to 'scale' to capture data variance.\\
\emph{MLP}: ReLU activation enhanced feature learning in a 3-layer Multi-Layer Perceptron (MLP) classifier with 256, 64, and 32 neurons. Using line search, learning rate of 0.001 has used wit ADAM optimized model for efficient convergence. To avoid overfitting, dropout regularization at 0.25 improved classification performance.\\
\emph{Logistic Regression}: Logistic Regression model performed best with L2 regularization and a regularization strength (C) of 5. The 'lbfgs' solver was used for efficient convergence, and the model required 180 iterations to achieve optimal results.\\
\emph{Decision Tree}: Decision Tree classifier was optimized with a max\_depth of 15 to avoid overfitting. The model used 3 min\_samples\_split and 2 min\_samples\_leaf to control tree complexity. The Gini criterion was employed to determine the best splits.\\
Table \ref{Model performance comparison} compares binary and multi-class classification metrics of best performing models on PTB and MITBIH-AD datasets. Comparatively, the proposed framework performs well in NSR Vs MCI classification. Because of the non-stationary similarities between MCI and non-MCI, miscalssification happened among these classes, in the other classification tasks. The proposed \v{C}ech complex framework provided consistent performance with stabilized standard deviation on 12-lead and 2-lead datasets with varied sampling frequencies and specifications.\\ Table \ref{feature performance comparison} describes the efficiency of the proposed homological features performance and complexity in comparison with the existing methods. The approaches utilizing wavelet transform and time-frequency characteristics demonstrate reduced computational cost with minimal performance. Raw ECG signals processed with U-net, CNN, ECG-SMART-NET, and RNN based frameworks yield modest performance accompanied by a considerable cost. Furthermore, the proposed \v{C}ech complex framework, incorporating persistent homological properties, contended with other complex models and exhibited 2.8\% mean gain in identification task. It also demonstrates proficiency in multi-class categorization with encouraging metric values. \v{C}ech complex framework class prediction tasks are evaluated for statistical significance using a paired t-test ($p<0.02$). 
\vspace{-5mm}
\section{Conclusion}
In this letter, for the first time, a novel \v{C}ech complex approach is proposed to utilize the persistent homological features for classifying NSR, MCI, and non-MCI subjects using ECG signals. The proposed framework captures the connectivity among the intra-sessional trials using homological groups evolved in the generation of \v{C}ech complex. Further, the homotopy equivalence checking condition nullifies the impact of outliers in the simplices generation while preserving their significance in feature extraction. The performance of ML models demonstrates the efficacy of homological features in both binary and multi-class classification. Further, this TDA approach can be extended with techniques like two-dimensional warping and shallow network schemes for more effective comprehension of the complexities between MCI and non-MCI ECG trials \cite{c1} \cite{c2}.

\end{document}